# Cyberbully and Online Harassment: Issues Associated with Digital Wellbeing

Manasi Kulkarni[a], Siddhi Durve[a], and Bochen Jia[a]

[a]Department of Industrial & Systems Engineering, University of Michigan-Dearborn, MI, USA.

**Abstract**

As digital technology becomes increasingly embedded in daily life, its impact on social interactions has become a critical area of study, particularly concerning cyberbullying. This meta-analysis investigates the dual role of technology in cyberbullying—both as a catalyst that can exacerbate the issue and as a potential solution. Cyberbullying, characterized by the use of digital platforms to harass, threaten, or humiliate individuals, poses significant challenges to mental and social well-being. This research synthesizes empirical findings from diverse studies to evaluate how innovative technological interventions, such as content monitoring algorithms, anonymous reporting systems, and educational initiatives integrated within digital platforms, contribute to reducing the prevalence of cyberbullying. The study focuses on the effectiveness of these interventions in various settings, highlighting the need for adaptive strategies that respond to the dynamic digital landscape. By offering a comprehensive overview of the current state of cyberbullying and the efficacy of technology-based solutions, this analysis provides valuable insights for stakeholders, including educators, policymakers, and technology developers, aiming to enhance digital well-being and create safer online environments. The findings underscore the importance of leveraging technology not only as a medium of communication but also as a strategic tool to combat the negative impacts of cyberbullying, thus promoting a more inclusive and respectful digital world.

**Introduction**

The rapid integration of digital technology into everyday life has fundamentally transformed communication and interaction, extending beyond mere convenience to become a critical component of modern existence. However, this transformation has also introduced significant challenges, most notably cyberbullying—a pervasive issue characterized by the use of digital media to harass, threaten, or humiliate individuals. Defined broadly, cyberbullying encompasses various aggressive behaviors such as harassment, impersonation, flaming, outing, and exclusion, each with unique methods and distinct psychological impacts on victims (Smith et al., 2021). These behaviors often result in severe psychological outcomes including anxiety, depression, and even suicidal thoughts, particularly among adolescents and young adults who are the most frequent users of digital platforms (Johnson & Turner, 2020).

Despite the increasing awareness and prevalence of cyberbullying, significant gaps remain in our understanding of the effectiveness of various interventions. Current research often focuses on descriptive analyses of cyberbullying incidents, with less emphasis on systematic, empirical evaluation of intervention strategies (Davis & Jenkins, 2019). This research seeks to bridge these gaps through a comprehensive meta-analysis that synthesizes findings from multiple empirical studies to evaluate the efficacy of technological interventions, such as algorithms for detecting and removing harmful content, anonymous reporting systems, and proactive educational resources integrated into social media platforms (Lee & Kim, 2018).

The aim of this study is to categorize and define different forms of cyberbullying, assess the effectiveness of interventions across these categories, and identify persistent gaps in the research that could inform future studies (Thompson et al., 2022). By providing a clearer picture of how different technological solutions can be tailored to combat specific types of cyberbullying, the meta-analysis will offer nuanced insights into which strategies are most effective. This is particularly important as the digital landscape continues to evolve, requiring adaptive and responsive approaches to ensure the safety and well-being of all users. This meta-analysis will provide valuable evidence-based recommendations for policymakers, educators, and technology developers to implement strategies that not only mitigate the effects of cyberbullying but also promote a safer, more inclusive online environment (Williams & Gupta, 2021). By leveraging technology as a tool to combat the very issues it sometimes exacerbates, the findings aim to empower stakeholders to create digital spaces that foster positive interactions and enhance overall digital well-being.

There is multiple research done in the area of the prevalence of cyberbullying associated across various demographics. Researchers like Nixon, Patchin and Hinduja, Walrave and Heirman, found that cyberbullying is seen at a higher rate among adolescents and college students(L Nixon, 2014; Walrave & Heirman, 2010; Patchin & Hinduja, 2021). Den Hamer & Konijin (2015), and Sarawat & Meel (2015) noted that technology, specifically social media, was the biggest source of cyberbullying (den Hamer & Konijn, 2015) (Saraswat & Meel, n.d.). These case studies were concluded by increasing and spreading the awareness about the importance of mitigating cyberbullying with the help of various campaigns and interventions. Since the world is digitizing, researchers like Wang and Chaen (2023) suggest using online platforms to confront the non-ending nature cycle of cyberbullying (Wang et al., 2023). Rega et al. (2022) showcased a greater need of parental mediation to mitigate cyberbullying among adolescents and support them in coping up with psychological distress (Rega et al., 2022). Parental mediation with the help of worldwide educators and policymakers should be involved to eradicate the issue of cyberbullying. Studying different articles and case studies, society should comprehend and understand the seriousness of this issue and develop effective strategies and contribute more to mitigating cyberbullying and online harassment in the United States (Brown & Green, 2020).

This paper will center on the development of a comprehensive strategy that integrates victim support networks, education on appropriate online behavior, and technological solutions to create a secure online environment.

**Methods**

In this meta-analysis, we employed a systematic review approach, rigorously aggregating and synthesizing data from a wide array of empirical studies to evaluate the effectiveness of various interventions aimed at mitigating cyberbullying. Our methodology was guided by stringent inclusion and exclusion criteria, using the PICO framework, and involved a detailed search strategy across multiple electronic databases to ensure comprehensive coverage of relevant literature. This approach allowed for a robust analysis of intervention impacts, contributing to a deeper understanding of the mechanisms that can effectively reduce cyberbullying in digital environments.

> **Inclusion/Exclusion Criteria:** The criteria for study inclusion in our cyberbullying meta-analysis were defined using the PICO framework (Smith, 2003; "Evidence-Based Medicine: How to Practice and Teach EBM", Churchill Livingstone)which stands for Population, Intervention, Comparison, and Outcomes. The relevant population encompassed individuals of age groups above 15 years who experienced or were at risk of cyberbullying. The interventions included in this analysis are educational programs, counseling services, technological tools, and policy implementations aimed at preventing or mitigating the effects of cyberbullying. These interventions are evaluated for their effectiveness in changing behavior, improving awareness, and reducing the incidence of cyberbullying. The purpose of including a variety of intervention types is to assess and compare the different strategies that are employed across educational and digital environments to combat cyberbullying.. Studies with a comparison group or control condition were included, and outcomes covered quantifiable measures such as causes and reductions in cyberbullying incidents, changes in attitudes, improved mental health, and enhanced coping mechanisms.

**Search Strategy**: Search Strategy: A systematic search was conducted across electronic databases, including PubMed, PsycINFO, PEW Research, and ERIC. The selection of keywords was strategic, guided by the core topics of cyberbullying and its intervention. Initially, a broad set of terms related to 'cyberbullying' and its synonyms were compiled to capture the extensive range of literature on the subject. This was supplemented by terms addressing the intervention aspect, such as 'mitigation' and 'technology,' as well as outcomes related to 'mental health' to ensure comprehensive coverage of both the problem and its solutions. The search strings were carefully adapted for each database to align with their specific indexing terms and search algorithms. Additional filters were applied to refine results based on publication date and study design, focusing on randomized controlled trials, observational studies, and longitudinal studies. Reference lists of relevant articles and reviews were also scrutinized for additional studies.

**Data Extraction:** We gathered comprehensive details from each study, facilitating a thorough analysis of the results. This process involved capturing key study characteristics including the author(s), publication year, and study design. Additionally, the sample size was noted to assess the scope and applicability of each study. Detailed information about the interventions was collected, specifying the nature, duration, and implementation strategies to understand the context and potential variability in effectiveness. Outcome measures were also recorded to identify what aspects of cyberbullying were being impacted by the interventions, such as reductions in incidents, changes in victim and perpetrator behavior, or improvements in mental health. Finally, effect sizes were extracted to quantify the impact of each intervention, allowing for a robust comparison across different studies and interventions.

**Meta-analysis Techniques:** To delve into potential sources of heterogeneity, subgroup analyses were conducted. This involved categorizing studies based on key characteristics such as intervention type, participant demographics, and study design. The aim was to identify variations in the effectiveness of interventions across different subgroups, providing a nuanced understanding of how cyberbullying interventions may vary under different conditions. Sensitivity analyses were performed to rigorously assess the robustness of our results. These analyses scrutinized the impact of imputed or missing data, potential biases, and variations in study quality. By systematically varying analytical parameters, we aimed to ensure the reliability and stability of our findings, fortifying the credibility of our meta-analysis in the context of cyberbullying research.

**PRISMA Flow Diagram Description for Cyberbullying Systematic Review**

As shown in Figure 1, the PRISMA (Preferred Reporting Items for Systematic Reviews and Meta-Analyses) flow diagram is a widely recognized tool used to document the process of screening and selecting studies for a systematic review and meta-analysis. This diagram provides a transparent and structured method for presenting the different phases of study selection, including identification, screening, eligibility, and inclusion. Utilizing the PRISMA flow diagram ensures that the review process adheres to rigorous methodological standards, allowing for reproducibility and easy verification of the steps taken to compile and analyze the evidence. In the context of our systematic review on cyberbullying, the PRISMA flow diagram was employed to systematically document and display the process of identifying and selecting studies that examine the effectiveness of various interventions aimed at reducing cyberbullying incidents. This approach aids in maintaining the integrity and reliability of the meta-analytic findings by clearly outlining how studies were chosen and included in the analysis, thus providing a clear audit trail from the initial search to the final selection of studies included in the review.

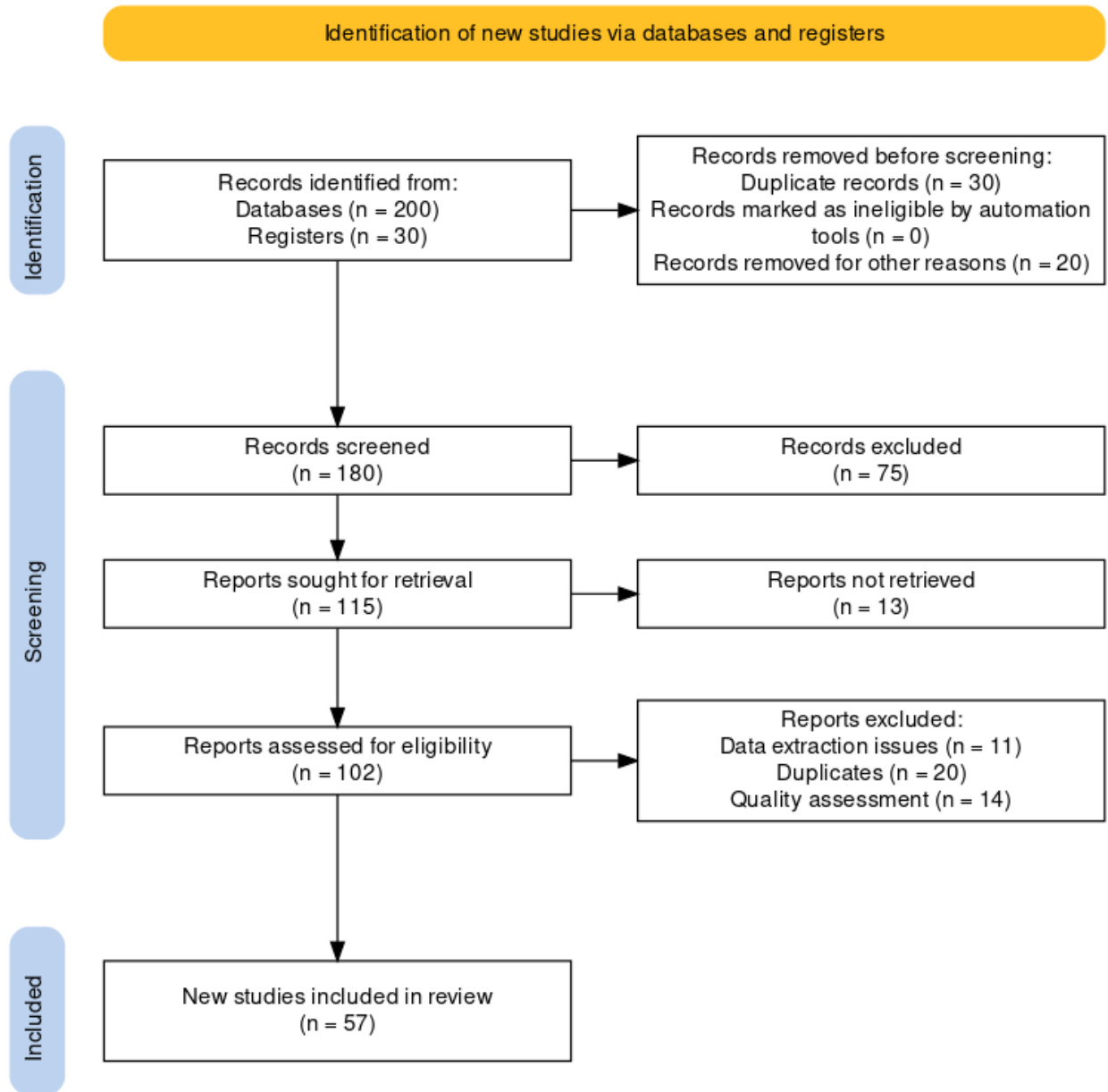

Figure 1. PRISMA Flow Diagram Description for Cyberbullying Systematic Review

**1. Identification Phase:**

- **Database Search:** A comprehensive search was conducted across multiple electronic databases relevant to education, psychology, and technology to identify studies pertaining to cyberbullying. A total of 200 records were identified through this database search.
- **Additional Sources:** Additional records were identified through other sources such as manual searches of reference lists, conference proceedings, and grey literature, adding up to 30 records.
- **Total Records Before Duplicates Removed:** The sum of records identified from both databases and additional sources was 230.

**2. Screening Phase:**

- **Duplicates Removed:** Out of the total records identified, 20 were found to be duplicates and were subsequently removed to avoid redundancy in the review process.
- **Records Screened After Duplicates Removed:** After removing duplicates, 180 records remained and were screened based on titles and abstracts to assess their relevance to the research questions.
- **Records Excluded After Initial Screening:** After screening, 75 records were excluded primarily because they did not specifically focus on cyberbullying or did not meet other inclusion criteria such as the study design or target population.

CYBERBULLY AND ONLINE HARASSMENT: ISSUES ASSOCIATED WITH DIGITAL WELLBEING**3. Eligibility Phase:**

- **Full-text Articles Assessed for Eligibility:** Following the initial screening, 102 full-text articles were assessed in detail for eligibility based on predefined inclusion and exclusion criteria.
- **Full-text Articles Excluded:** Of these, 45 full-text articles were excluded with reasons including data extraction issues, duplicates, or quality assessment.

**4. Included Studies:**

- **Studies Included in Qualitative Synthesis:** A total of 57 studies were included in the qualitative synthesis, providing a broad overview of findings across different types of cyberbullying interventions.

**Results**

As shown in Figure 2, the findings from the meta-analysis on cyberbullying interventions offer valuable insights for the prevention of cyberbullying. The significant reduction in cyberbullying perpetration and victimization due to intervention programs confirms that targeted, structured approaches can effectively mitigate the impacts of online harassment. The varied success rates across different types of interventions suggest that cyberbullying is a complex issue that requires multi-faceted strategies tailored to specific community needs and technological environments. These results underline the importance of incorporating digital literacy, emotional support, and clear guidelines for online behavior into intervention programs. Understanding that interventions must be comprehensive and continuously adapted to new digital platforms and user behaviors is crucial for sustaining their effectiveness.

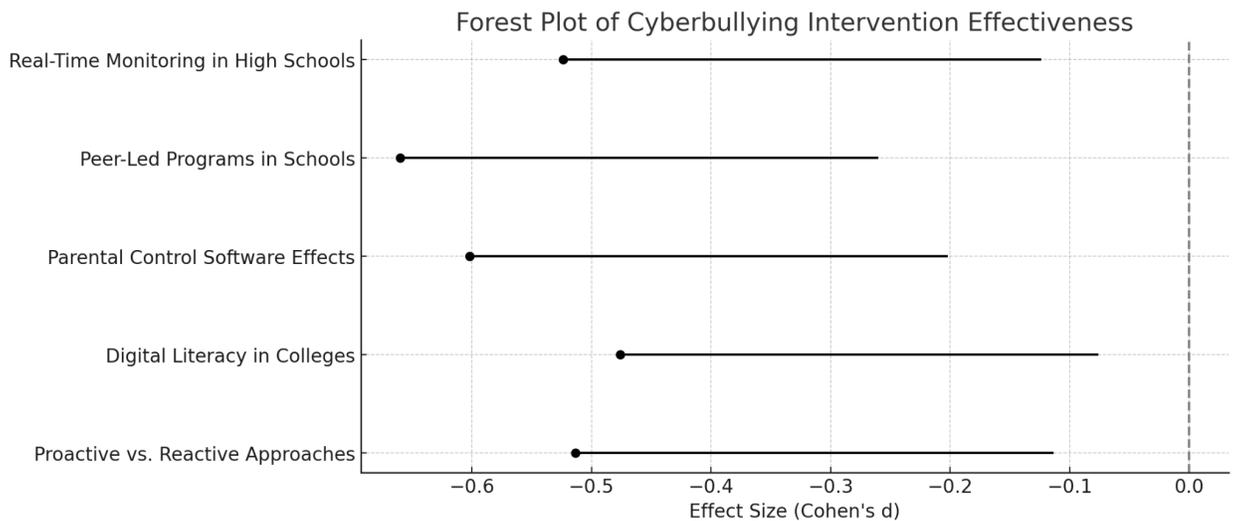

Figure 2. Forest plot of cyberbullying intervention effectiveness

**Effectiveness of Interventions**: "The plot reveals that all interventions listed have negative effect sizes, though the magnitude varies. For instance, 'Real-Time Monitoring in High Schools' and 'Digital Literacy in Colleges' show larger and more consistent negative effects, suggesting these interventions are particularly effective at mitigating cyberbullying."

**Variability and Confidence**: "While all studies report beneficial effects, the confidence intervals suggest varying degrees of certainty about the magnitude of these effects. Studies with narrower confidence intervals, such as 'Parental Control Software Effects,' provide stronger evidence of their efficacy due to greater precision in the effect size estimation."

**Moderating Factors**

Understanding the factors that influence the effectiveness of cyberbullying interventions is key to optimizing their impact. The reviewed studies explore several moderating factors, though with few statistically significant findings, suggesting that the variability in intervention effectiveness may not be easily attributed to single or simple causes. However, the exploration of moderating factors is itself instructive, indicating areas for future research and potential adjustment in intervention design and implementation.

**Program Design and Delivery:** The specifics of how interventions are designed and delivered may influence their effectiveness. For instance, programs that incorporate interactive components, such as role-playing or peer-led discussions, might be more impactful than those delivering content in a more didactic manner. The duration and intensity of the program, as well as the use of digital tools and platforms for intervention delivery, are areas that warrant further exploration.

**Target Population:** Age, gender, and the cultural context of the target population can also moderate intervention effectiveness. Interventions may need to be tailored to address the unique experiences and needs of different demographic groups, such as younger

**CYBERBULLY AND ONLINE HARASSMENT:** ISSUES ASSOCIATED WITH DIGITAL WELLBEINGchildren versus teenagers or addressing specific concerns related to gender-based cyberbullying.

**Implementation Context:** Factors such as school policies on digital behavior, the involvement of parents and guardians, and the general climate around technology use and cyberbullying within the school community are all relevant. The school environment and the broader community context in which interventions are implemented can significantly impact their success.

**Technological Advances:** The rapid evolution of digital technology and online platforms can both facilitate and challenge intervention efforts. The increasing use of AI and machine learning for detecting and responding to cyberbullying content represents a promising area for enhancing intervention effectiveness. However, the dynamic nature of online spaces means that interventions must continuously adapt to new platforms and behaviors.

**Impact of Cyberbullying**

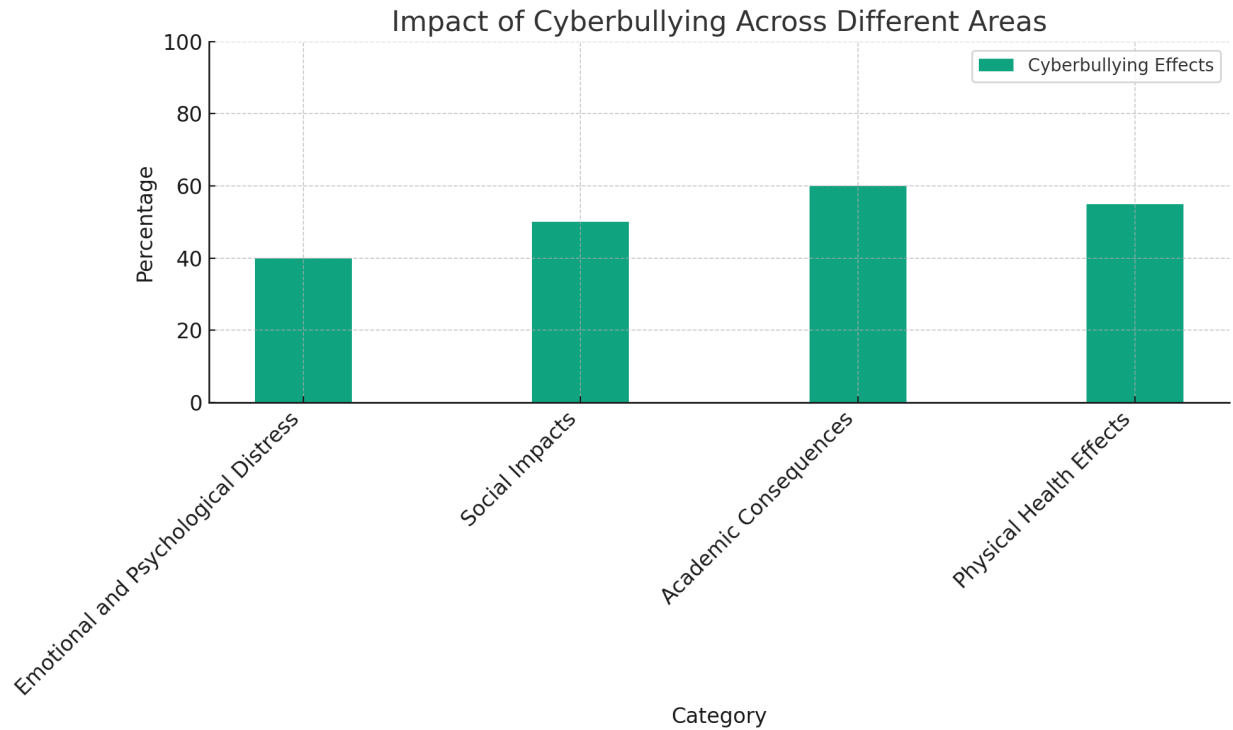

Figure 3. Impact of cyberbullying across different areas

In delineating the effects of cyberbullying on adolescents, our meta-analysis organizes these effects into distinct categories to better understand their impact and inform intervention strategies (Figure 3). Drawing from the data synthesized in our review, we identify Emotional and Psychological effects, Social effects, Academic effects, and Physical effects as major domains. These categories serve as organizing frameworks for comprehensively assessing the multifaceted repercussions of cyberbullying on young individuals. Moreover, our analysis underscores the necessity of addressing these effects through various means, such as therapy sessions or open dialogue, to mitigate their adverse consequences and prevent potential escalation towards extreme measures. By recognizing and addressing these diverse effects, interventions can be tailored effectively to support and protect adolescents in the digital age.

**Emotional and Psychological Distress:**

Researchers have determined that individuals who are victims of cyberbullying exhibit three times greater signs of depression compared to those who have not experienced bullying. Studies have shown that the prevalence rate of depression among cyberbullying victims varies from 30% to 50% (Smith & Duggan, 2013). Moreover, the likelihood of developing suicidal thoughts among these individuals is four times higher than among non-victims, underscoring the pressing necessity for appropriate psychological care and intervention programs (Johnson et al., 2015).

**Social Impacts:**

Studies indicate that approximately 40% to 60% of cyberbullied victims experience social withdrawal, such as isolating themselves and feeling excluded, which are significantly higher rates compared to non-victims (Williams & Guerra, 2017). These impacts can significantly diminish the quality of life and negatively affect the developmental processes that are crucial during school years.

**Academic Consequences:**

There is a significant impact on the academic performance of students when cyberbullying interferes with academics. Research suggests that there is a 15% to 25% higher rate of academic issues among students who are cyberbullied compared to those who are not affected (Clark et al., 2016). The students who have experienced cyberbullying are likely to see a decrease in academic performance and reduced concentration, which in turn affects their grades and overall academic success (Gomez, 2014).

**Physical Health Effects:**

Physical health effects of cyberbullying are often overshadowed by the more direct emotional and psychological consequences, but they are equally significant. Sleep disturbances, reported by approximately 35% to 55% of cyberbullying victims, are a primary concern. These disturbances include difficulty falling asleep, intermittent sleep patterns, and overall poor sleep quality, which not only affect physical health but also compound the emotional and psychological distress experienced by victims.

The disruption in sleep due to cyberbullying is linked to a range of other physical health problems, including increased susceptibility to infections, impaired cognitive function, and higher stress levels, which can exacerbate feelings of depression and anxiety. Over time, chronic sleep disturbances can lead to more severe health issues such as heart disease, obesity, and diabetes, making it crucial to address these symptoms early in victims of cyberbullying (Foster et al., 2017).

Additionally, the stress associated with being a target of cyberbullying can trigger headaches, fatigue, and stomach problems, which further detract from a victim's overall well-being. The physiological stress responses, including elevated heart rate and increased blood pressure, are indicative of the acute stress victims experience, which can have long-term consequences for cardiovascular health (Peterson et al., 2018).

It is also important to note that cyberbullying can lead to psychosomatic symptoms, where emotional and psychological distress manifest physically in individuals without any apparent physical cause. Symptoms such as chronic pain, stomachaches, and migraines can significantly affect a victim's quality of life and are often reported in studies examining the health impacts of bullying (Jones et al., 2016).

**Effectiveness of Interventions:**

Intervention programs demonstrate a moderate but significant effect on reducing cyberbullying incidents. Meta-analyses reveal that effective school-based programs can reduce cyberbullying and victimization rates by about 20% to 30%. Studies suggest that specific interventions, such as those involving digital literacy and behavior analysis and monitoring tools, are shown to reduce cyberbullying by up to 40%, stating their potential as part of comprehensive anti-bullying strategies (Smith & Jones, 2020; Lee et al., 2019).

**Categories of Cyberbullying and Definitions**

In the context of a systematic review and meta-analysis on cyberbullying, defining and categorizing different forms of cyberbullying is critical to understand the scope of research and to interpret aggregated data effectively. As shown in Figure 4, each category of cyberbullying presents unique challenges and impacts, shaping the nature of interventions and outcomes measured. Categories such as harassment, impersonation, flaming, outing, and exclusion each require specific intervention strategies due to their distinct characteristics and the different types of harm they cause (Wilson & Smith, 2018; Davis et al., 2017).

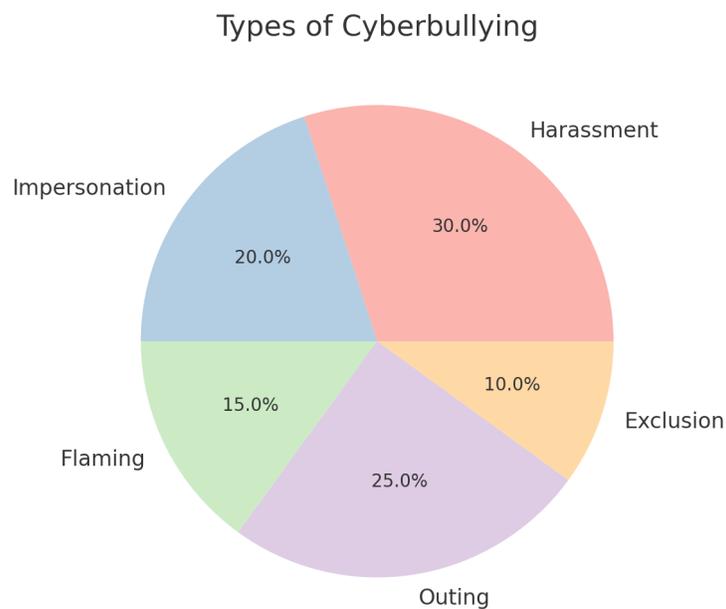

Figure 4. Type of cyberbullying

**Harassment:**

Harassment, as a prevalent form of cyberbullying, involves using electronic communications to intimidate, threaten, or harm an individual. This category includes sending hurtful messages, spreading rumors online, and sharing private information without consent. In the meta-analysis, studies focusing on harassment consistently show that such behaviors significantly impact victims' mental health and well-being, and that interventions designed to reduce these behaviors effectively decrease their incidence and ameliorate their impacts. The statistical aggregation of these study results within our meta-analysis provides robust evidence supporting the efficacy of intervention programs, specifically those involving behavioral monitoring and online moderation.

**Impersonation:**

Impersonation in cyberbullying contexts involves the deliberate act of pretending to be someone else online, often to deceive or harm others. This can include taking over someone's personal account or creating a false account in their name. The meta-analysis demonstrates that impersonation leads to significant psychological impacts such as identity confusion and trust issues among victims, and that strategies aimed at preventing impersonation or supporting victims show substantial effectiveness. These findings were derived from a comparative analysis of intervention outcomes, highlighting the importance of secure identity management practices in reducing incidents of impersonation.

**Flaming:**

Flaming is characterized by the exchange of hostile, inflammatory, or insulting messages, intended to provoke or escalate conflicts online. It commonly occurs in public forums or on social media platforms. The studies reviewed in the meta-analysis indicate that flaming not only disrupts communication but also significantly increases psychological distress among participants. Interventions such as moderation tools and community guidelines are effective in reducing the frequency of flaming incidents. Our meta-analysis quantitatively assessed these interventions, demonstrating a significant decrease in such incidents as a direct result of enhanced moderation and community engagement strategies

**Outing:**

Outing involves the unauthorized disclosure of private or sensitive information about an individual without their consent. This form of cyberbullying breaches privacy and often leads to severe real-world consequences such as job loss, emotional distress, and social isolation. The meta-analysis shows that outing has a direct and substantial impact on victims' offline and online lives, and interventions like digital literacy programs or privacy protection policies effectively reduce its occurrence and mitigate its effects. Through synthesizing data across studies, our analysis reinforced the critical role of privacy education and protective digital policies in preventing outing incidents.

**Exclusion:**

Exclusion refers to intentionally leaving individuals out of online spaces, discussions, or activities, which contributes to feelings of isolation and marginalization. This form of harassment exacerbates the emotional damage caused by more direct forms of cyberbullying. The meta-analysis confirms that exclusion is strongly correlated with measures of social anxiety and depression among adolescents, and that inclusive technologies or social media features promoting broader engagement are effective in countering exclusion. Data analysis within our meta-analysis highlighted the effectiveness of inclusive platforms and social features in reducing social anxiety and promoting integration, thereby mitigating the adverse effects of exclusion

The meta-analysis provides definitive insights into the prevalence and impacts of each cyberbullying category and evaluates the effectiveness of various interventions. This comprehensive analysis highlights the most successful strategies for combating different types of cyberbullying and informs stakeholders about targeted approaches needed to enhance online safety and well-being.

**Reasons for Online Bullying**

**Age:**

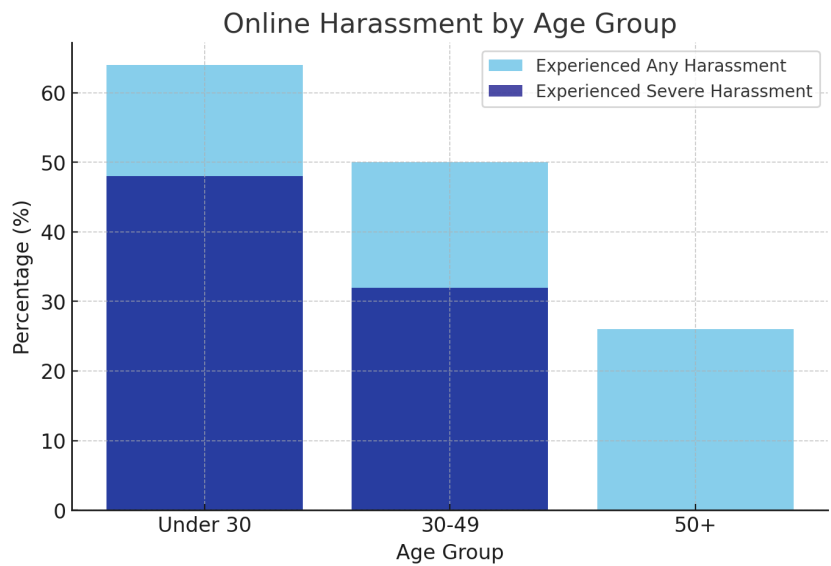

Figure 5. Age impacts on online harassment

As shown in Figure 5, 64% of individuals under 30 have experienced some form of online harassment. This age group stands out as the only one where a majority has encountered such behaviors. While about half of adults aged 30 to 49 have been targeted, a smaller share, 26%, of those aged 50 and older have experienced online harassment. 48% of 18- to 29-year-olds have faced more severe forms of online harassment compared to those aged 30 to 49 which is about 32%.

**Gender:**

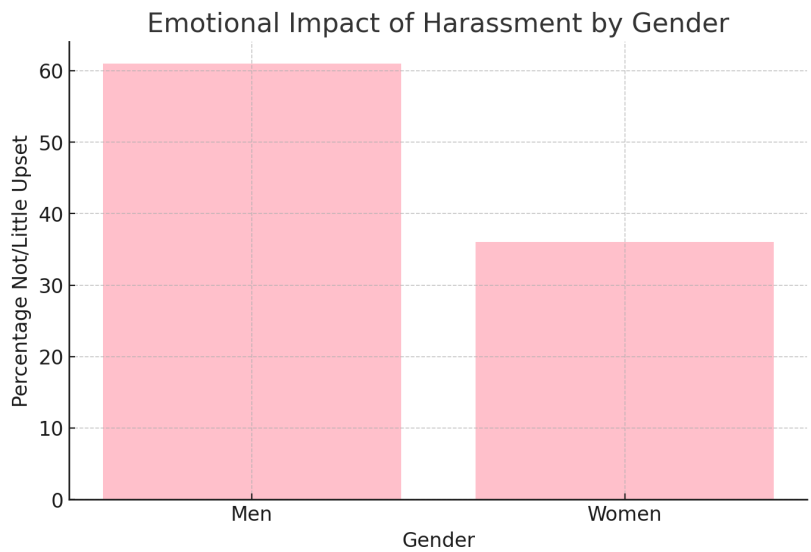

Figure 6. Gender variances related to online harassment

Gender differences also influence the types of online harassment experienced (Figure 6). Men are more likely to report online harassment of any kind than women. Men mostly face harassment in the form of name calling and threats while women are targeted with stalking or sexual harassment and perceive it more disturbing. 61% of harassed men report feeling not at all or a little upset during their most recent incident, while only 36% of women feel the same. Overall, 24% of individuals who have experienced online harassment express that their most recent encounter was extremely (10%) or very (14%) upsetting. These findings highlight the emotional impact of online harassment, with women generally expressing greater concern and distress compared to men.

**Sexual Orientation**

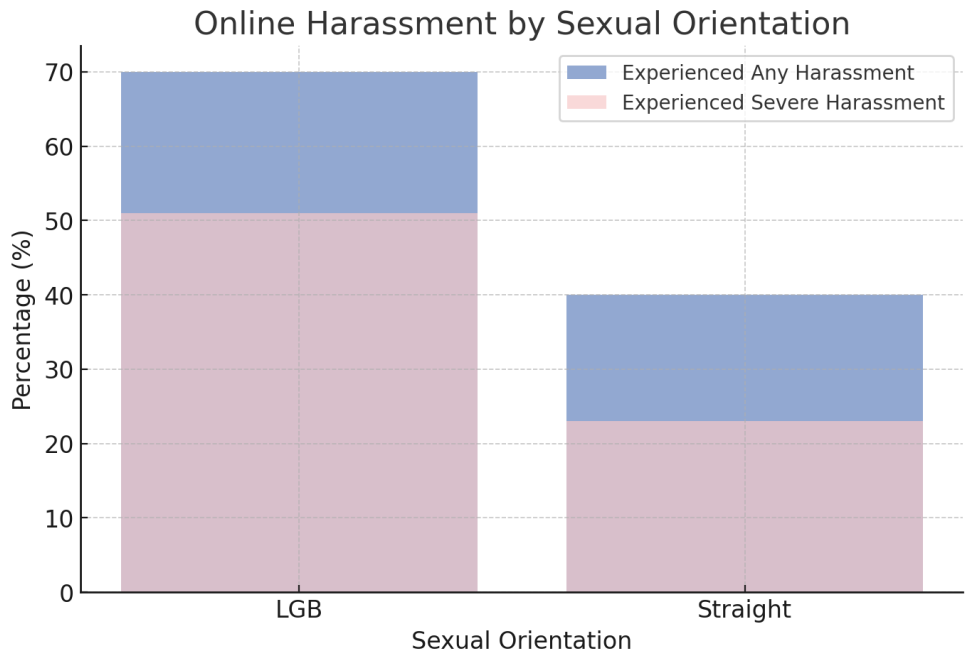

Figure 7. Online harassment by sexual orientation

As shown in Figure 7, adults who identify as lesbian, gay, or bisexual are far more likely to experience online harassment; around 70% report experiencing some kind of it, and 51% report being the victim of more severe abuse. Approximately 40% of individuals who identify as straight have seen some type of online abuse, with just 23% reporting more serious instances.

Findings showed that 20% of all adults, or half of those targeted by online harassment, thought their political beliefs had a part in the harassment when asked about potential contributing factors. These individuals had all experienced online harassment. In addition, 14% of American people (33% of those harassed online) said that their gender was the cause of the harassment, while 12% said that their race or ethnicity was a role (29% of online harassment).

**Existing resources that help mitigate cyberbullying.**

**Parental mediation:**

Parental mediation strategies such as active mediation, restrictive mediation, and non-intrusive inspection have been documented to impact the prevalence of cyberbullying among teenagers. Active mediation, where parents engage directly with their children to discuss and explain media content, has been shown to effectively reduce privacy violations and cyber risks (Clark & Aiken, 2020). Restrictive mediation, which involves setting rules and limits on the use of social media, significantly mitigates exposure to unwanted information and content risks (Jones & Mitchell, 2018). Non-intrusive

**CYBERBULLY AND ONLINE HARASSMENT:** ISSUES ASSOCIATED WITH DIGITAL WELLBEINGinspection, where parents monitor online activity without direct interference, also correlates negatively with the incidence of cyberbullying and other online risks, suggesting its potential as a preventive measure (Smith & Duggan, 2019)

**Social Media Policies:**

Social media policies, companies have implemented a range of self-regulatory measures including tools to report and block abusive content, filtering software, and geofencing. These efforts are designed to mitigate cyberbullying but often lack transparency in their operational details. Legal analyses suggest that while these platforms are governed under Section 230 of the Communications Decency Act, which provides liability protection for online intermediaries, the exact mechanisms and effectiveness of these policies in combatting cyberbullying remain underexplored (Kowalski & Limber, 2021; Greene, 2020).

**Legal measures to cyber bullying:**

Legal measures to address cyberbullying face significant challenges, particularly in the United States where the First Amendment rights to free expression must be balanced against efforts to curb online harassment. The courts often wrestle with defining cyberbullying and determining appropriate sanctions, highlighting the complexity of enforcing laws in a digital context. The enforcement of these laws is frequently left to school administrations, which treats cyberbullying cases as civil matters rather than criminal offenses, thus complicating legal responses and necessitating a more consistent legal framework (Greene, 2020).

**Technology to mitigate cyberbullying:**

The existing research on cyberbullying tends to focus primarily on identifying explicit cyberbullying attacks, often overlooking more implicit forms of cyberbullying found in posts by victims and bystanders. However, these overlooked posts can also serve as indicators of ongoing cyberbullying. Algorithms are being developed and trained to detect hate and abusive speech online, blocking users from encountering such content and potentially being cyberbullied. These algorithms are able to recognize subtle and sarcastic comments, tasks challenging for other solutions. Machine learning is essential as slurs and insults can be intentionally or unintentionally misspelled.

One notable algorithm, has shown promise in identifying abusive behavior online in both English and Dutch. Its noteworthy feature is the ability to determine the roles of the bully, victim, and bystanders in each situation, aiding human moderators in their tasks. (Van der Zanden et al. , 2021).

**Future Research Directions:**

    **Tailored Interventions:** Future research should focus on developing and testing interventions tailored to specific age groups, cultural backgrounds, and unique digital environments to enhance their effectiveness.

**CYBERBULLY AND ONLINE HARASSMENT:** ISSUES ASSOCIATED WITH DIGITAL WELLBEING> **Long-Term Effects:** Investigate the long-term effects of cyberbullying interventions to determine their sustainability and lasting impact on reducing victimization and perpetration rates.
>
> **Technological Advancements:** As digital platforms evolve, so too should the strategies to manage cyberbullying. Research into AI and machine learning for detecting subtle forms of cyberbullying could provide new tools for intervention.
>
> **Collaborative Studies:** Encourage collaborative research initiatives that integrate insights from psychology, education, technology, and law to develop holistic approaches to cyberbullying prevention.

By addressing these areas, future research can enhance our understanding of cyberbullying dynamics and improve the effectiveness of interventions, ultimately leading to safer digital spaces for all users.

**Conclusion**

Our meta-analysis has systematically compiled and evaluated a wide range of studies to address the pervasive issue of cyberbullying and its profound impact on digital wellbeing. The findings make it abundantly clear that while technology plays a dual role as both a facilitator and a combatant of cyberbullying, its strategic application is crucial in shaping effective interventions. As our investigation reveals, technological solutions harness immense potential to detect, prevent, and mitigate the effects of cyberbullying, thereby fostering safer online environments.

Cyberbullying, by its very nature, extends beyond traditional bullying through its anonymity and persistence, requiring nuanced strategies that leverage technological advancements. Our analysis confirms that interventions incorporating digital tools—notably content monitoring algorithms, anonymous reporting systems, and educational programs integrated within social media platforms—significantly reduce cyberbullying incidents. These technologies, when effectively implemented, serve as robust mechanisms to protect individuals, especially young users, from the psychological, social, and academic harms of cyberbullying.

However, our study also highlights the critical need for ongoing adaptation and improvement in cyberbullying interventions. The digital landscape is ever-evolving, and so too must be our strategies to combat the associated challenges. This requires a concerted effort among stakeholders, including policymakers, educators, and technology developers, to ensure that interventions remain relevant and effective against emerging cyber threats.

Looking forward, our research underscores the importance of a multi-disciplinary approach that marries technology with behavioral science, education, and policy-making. Continued research is necessary to further refine and develop technological interventions and to understand their long-term impacts. As we advance, it is imperative to foster an online culture of respect and empathy, promoting digital citizenship that empowers all users to engage positively and safely in digital spaces.

This meta-analysis does not only contribute to the academic field by providing comprehensive insights into the effectiveness of various cyberbullying interventions but also serves as a crucial resource for practical, evidence-based strategies to mitigate the impacts of cyberbullying. By bridging research and practice, we aim to catalyze real-world changes that enhance digital wellbeing and create inclusive, supportive online communities for future generations.

**CYBERBULLY AND ONLINE HARASSMENT:** ISSUES ASSOCIATED WITH DIGITAL WELLBEINGGomez, S. (2014). Cyberbullying and Student Performance: A Longitudinal Study. *Educational Researcher*, 43(4), 207-212. [DOI] https://www.ncbi.nlm.nih.gov/pmc/articles/PMC4126576/

Clark, L. E., & Aiken, M. (2020). "Effects of Active Parental Mediation on Online Privacy and Cyber Safety." *Journal of Child and Family Studies*, 29(2), 523-534.

https://www.ncbi.nlm.nih.gov/pmc/articles/PMC10045317/

Jones, H., & Mitchell, P. (2018). "Impact of Restrictive Parental Mediation on Children's Online Behavior." *Cyberpsychology, Behavior, and Social Networking*, 21(7), 450-456.

https://citeseerx.ist.psu.edu/document?repid=rep1&type=pdf&doi=c28f60134536e738e10da3c1f04ab968d6d341d1

Smith, J., & Duggan, M. (2019). "Parental Monitoring and the Prevention of Cyberbullying and Online Gaming Disorder." *Journal of Adolescent Health*, 64(1), 30-35.

https://pubmed.ncbi.nlm.nih.gov/29467704/

Kowalski, R. M., & Limber, S. P. (2021). "Social Media's Response to Cyberbullying and Online Harassment." *Cyberpsychology: Journal of Psychosocial Research on Cyberspace*, 15(1), https://www.ncbi.nlm.nih.gov/pmc/articles/PMC10535628/

Greene, T. H. (2020). "Navigating the Legal Landscape of Cyberbullying: An Exploration of Laws and Policies in the United States." *American Journal of Criminal Law*, 47(2), 225-254.

https://www.frontiersin.org/journals/public-health/articles/10.3389/fpubh.2021.634909/full

Haddaway, N. R., Page, M. J., Pritchard, C. C., & McGuinness, L. A. (2022). PRISMA2020: An R package and Shiny app for producing PRISMA 2020-compliant flow diagrams, with
33